%% file: main.tex
\documentclass[reprint,superscriptaddress,amsmath,amssymb,aps,prl,floatfix,longbibliography]{revtex4-2}

\usepackage{graphicx}
\usepackage{dcolumn}
\usepackage{bm}
\usepackage{hyperref}

\usepackage[dvipsnames]{xcolor}
\usepackage[normalem]{ulem}

\newcommand{\ilm}{Univ Lyon, Univ Claude Bernard Lyon 1, CNRS, Institut Lumi\`ere Mati\`ere, F-69622, VILLEURBANNE, France}
\newcommand{\iuf}{Institut Universitaire de France (IUF), 1 rue Descartes, 75005 Paris, France}

\begin{document}


\title{Microscopic origins of the viscosity of a Lennard-Jones liquid}

\author{Farid Rizk}
\affiliation{\ilm}
\author{Simon Gelin}
\affiliation{\ilm}
\author{Anne-Laure Biance}
\affiliation{\ilm}
\author{Laurent Joly}
\email{laurent.joly@univ-lyon1.fr}
\affiliation{\ilm}
\affiliation{\iuf}

\date{\today}

\begin{abstract}
Unlike crystalline solids or ideal gases, transport properties remain difficult to describe from a microscopic point of view in liquids, whose dynamics result from complex energetic and entropic contributions at the atomic scale. Two scenarios are generally proposed: one represents the dynamics in a fluid as a series of energy barrier crossings, leading to Arrhenius-like laws, while the other assumes that atoms rearrange themselves by collisions, as exemplified by the free volume model. To assess the validity of these two views, we computed, using molecular dynamics simulations, the transport properties of the Lennard-Jones fluid and tested to what extent the Arrhenius equation and the free volume model describe the temperature dependence of the viscosity and of the diffusion coefficient at fixed pressure. Although both models reproduce the simulation results over a wide range of pressure and temperature covering the liquid and supercritical states of the Lennard-Jones fluid, we found that the parameters of the free volume model can be estimated directly from local structural parameters, also obtained in the simulations. This consistency of the results gives more credibility to the free volume description of transport properties in liquids.
\end{abstract}

\maketitle

\paragraph*{Introduction--}
Modeling how liquids flow is a subject of great fundamental interest and of major importance in many industrial applications (composite molding \cite{Henne2004}, lubricants \cite{Quinchia2010}, pharmaceutical \cite{M.TakechiC.Uno1994}, etc.). The main property that characterizes the flow of liquids is viscosity, which depends on the nature of the liquid and its environment, in particular pressure and temperature. Experimental research work has shown, since the 1930s, that the temperature dependence of viscosity obeys an Arrhenius-type equation, called the Andrade's law, in a wide variety of liquids \cite{andrade1930viscosity,DaviesMatheson1967,Poirier1988,MessaadiDhouibiHamdaBelgacemAdbelkaderOuerfelliHamzaoui2015}. Drawing on this observation, some models based on Eyring's equation propose that viscosity is controlled by the breaking of interatomic bonds in the liquid (i.e. the crossing of energy barriers) \cite{Eyring1936a,MaciasSalinasDuranValenciaLopezRamirezBouchot2008,HeyesDiniSmith2018}, but the identification of these breaking events is questionable in simple liquids. We will call this approach \textit{Eyring's model}.
Another line of thought, embodied by the free volume model, describes the dynamics of atoms as a succession of hard-sphere-like collisions that may result in local mass transport. Although the free volume model has been used to describe the transport properties of many liquids \cite{Cohen1959, Turnbull1961, Turnbull1970, Falk2020}, it is based on a microscopic concept -- the free volume -- that is difficult to relate to local atomistic parameters. Moreover, being derived at constant density, this model does not offer a direct explanation of the experimentally observed Arrhenius law.

In this work, we use molecular dynamics (MD) simulations to measure the viscosity of the one-particle Lennard-Jones (LJ) fluid and discriminate between these two descriptions of transport. The LJ model, in addition to exhibiting the generic structural and dynamical properties of most liquids bound by van der Waals or metallic interactions~\cite{Dyre2014}, has been shown to quantitatively reproduce the behavior of various atomic or molecular fluids, such as rare-gas liquids, carbon dioxide, and linear or aromatic hydrocarbons \cite{Liu1998,Galliero2005}. 
We start by measuring viscosity across the LJ fluid's pressure-temperature phase diagram and then determine the region in which the Arrhenius equation holds. As the two considered models are intrinsically built to describe diffusion rather than viscosity, we also evaluate diffusion coefficients and show that the Stokes-Einstein law holds in the Arrhenius-viscosity domain. This enables us to analyze the microscopic origin of the parameters entering in the description. 
Although both Eyring and free volume models contain two parameters, we find that one of the parameters of the free volume model -- the intrinsic particle volume -- can be evaluated directly from the simulations, by analyzing the fluid's microstructure. 
Eventually, we discuss the parameters of the Arrhenius equation in light of the free volume model and suggest that the emergence of the Arrhenius law, although compatible with an energetic picture, is only accidental.

\paragraph*{Methods--}
We carry out molecular dynamics simulations, using the LAMMPS code \cite{LAMMPS2022}, to compute the viscosity and diffusion coefficient of a one atom type 12-6 LJ fluid, defined by the following interaction potential: 
$U(r)=4\epsilon\left[\left(\frac{\sigma}{r}\right)^{12}-\left(\frac{\sigma}{r}\right)^{6}\right]$ , 
where $\epsilon$ and $\sigma$ are the usual potential energy and particle diameter parameters, respectively, and $r$ is the interatomic distance. The potential is truncated 
at a cutoff distance $r_c=3\sigma$. 
The simulations are run at constant number of particles, temperature and pressure, and cover a wide range of temperature and pressure values in the liquid and supercritical 
phases. Details about values explored here are given in the supplemental material (SM) \cite{sm}.
All of our results are reported in LJ reduced units, using $\epsilon$, $\sigma$, and the particle mass $m$ as measures of energy, distance, and mass.

We followed the Green-Kubo method to compute the viscosity $\eta$ \cite{allen2017computer,Heyes2019}:
$\eta =\lim_{t  \rightarrow \infty}\frac{V }{T } \int_{0}^{t} C_{ij}(\tau) \mathrm{d}\tau$, 
with $V$ the simulation box volume, $T$ the temperature, $t$ the final correlation time and $C_{ij}(\tau)$ the auto-correlation function, at time $\tau$, of the non-diagonal elements $p_{ij}$ of the pressure tensor: $C_{ij}(\tau)=\langle p_{ij}(\tau)\, p_{ij}(0) \rangle$.
We computed the diffusion coefficient $D_1$ from the mean-square displacement $\langle r^2(t) \rangle$ in the diffusive regime, as follows:
$D_1=\frac{1}{6} \frac{\mathrm{d}}{\mathrm{d}t}\left<{r}^2\right> $. To take into account the hydrodynamic interactions between the periodic image boxes, we made the following correction to the diffusion coefficient: $D=D_1+2.837\frac{T}{6\pi \eta L_{box}}$, with $L_{box}$ the size of the cubic simulation box which we calculated as $L_{box}=(250/\rho)^{1/3}$.
More details on the MD simulations can be found in the SM \cite{sm}.

\paragraph*{Results--}

\begin{figure}
    \centering
    \includegraphics[width=\linewidth]{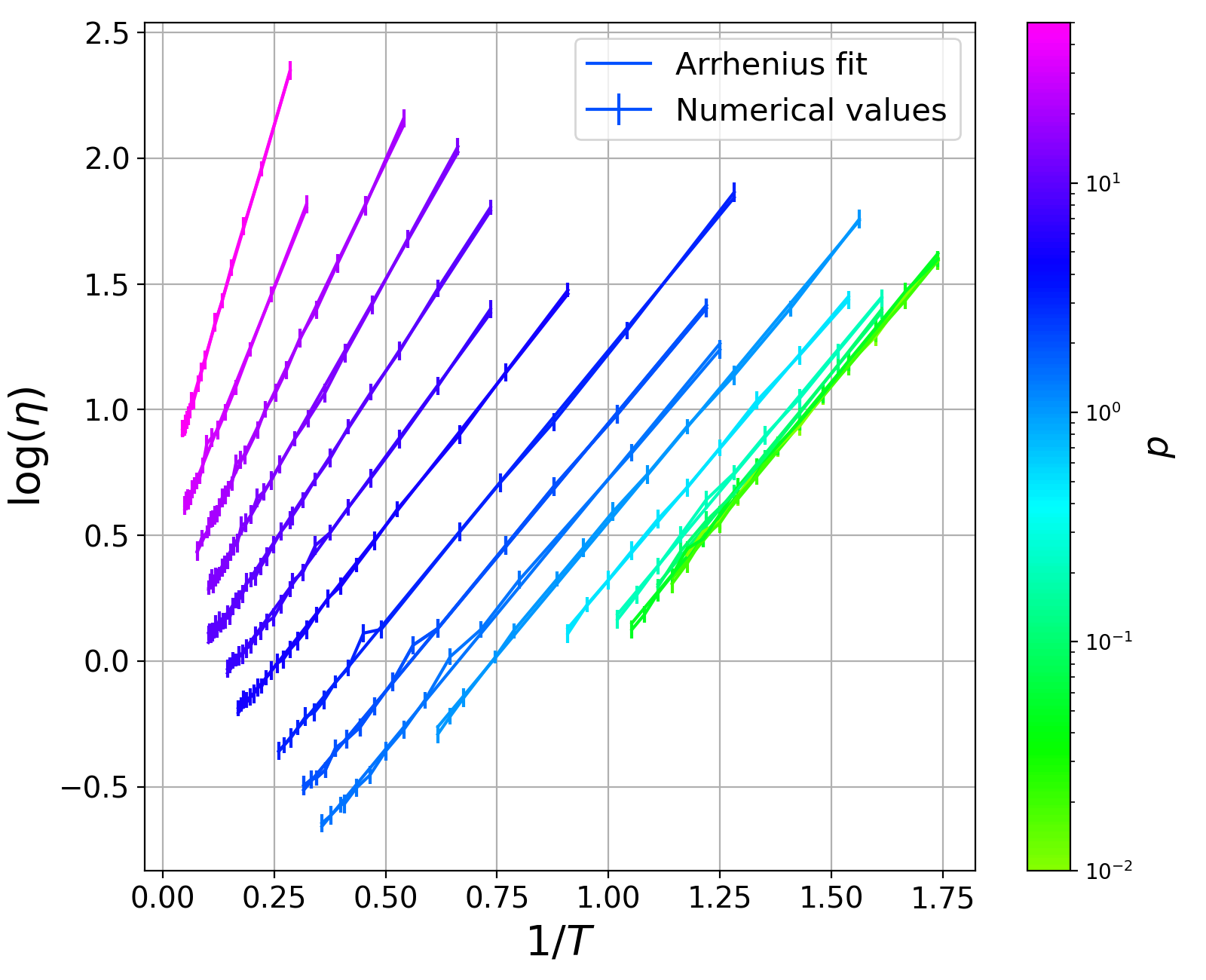}
    \caption{Arrhenius plot of the temperature dependence of viscosity for pressures varying from $p=0.01$ to $p=50$. At each pressure, simulation results are represented in a range of temperatures over which they are well fitted by an Arrhenius law (straight line). 
    }
    \label{fig:visc_temp_pres}
\end{figure}

After having computed the viscosity $\eta$ at different temperatures and pressures in the liquid and supercritical 
phases, we fitted the temperature dependence of $\eta$, at fixed pressure, with the Arrhenius law \cite{Carvalho-Silva2019}:
\begin{equation}\label{eq:arrh_visc}
\eta(T;p)=\eta_0(p)\, e^{\frac{Q(p)}{T}} , 
\end{equation}
where $\eta_0$ and $Q$ are the pre-exponential factor and the activation energy, respectively.
The results reported in Fig.~\ref{fig:visc_temp_pres} show that there exists, at each pressure, a significant range of temperatures over which viscosity is well described by the Arrhenius model. 
As shown in the SM \cite{sm}, the activation energy $Q$ is constant with pressure at low pressures, and increases for $p\gtrsim 3$; in contrast, $\eta_0$ increases at all pressures.

\begin{figure}
    \centering
    \includegraphics[width=0.9\linewidth]{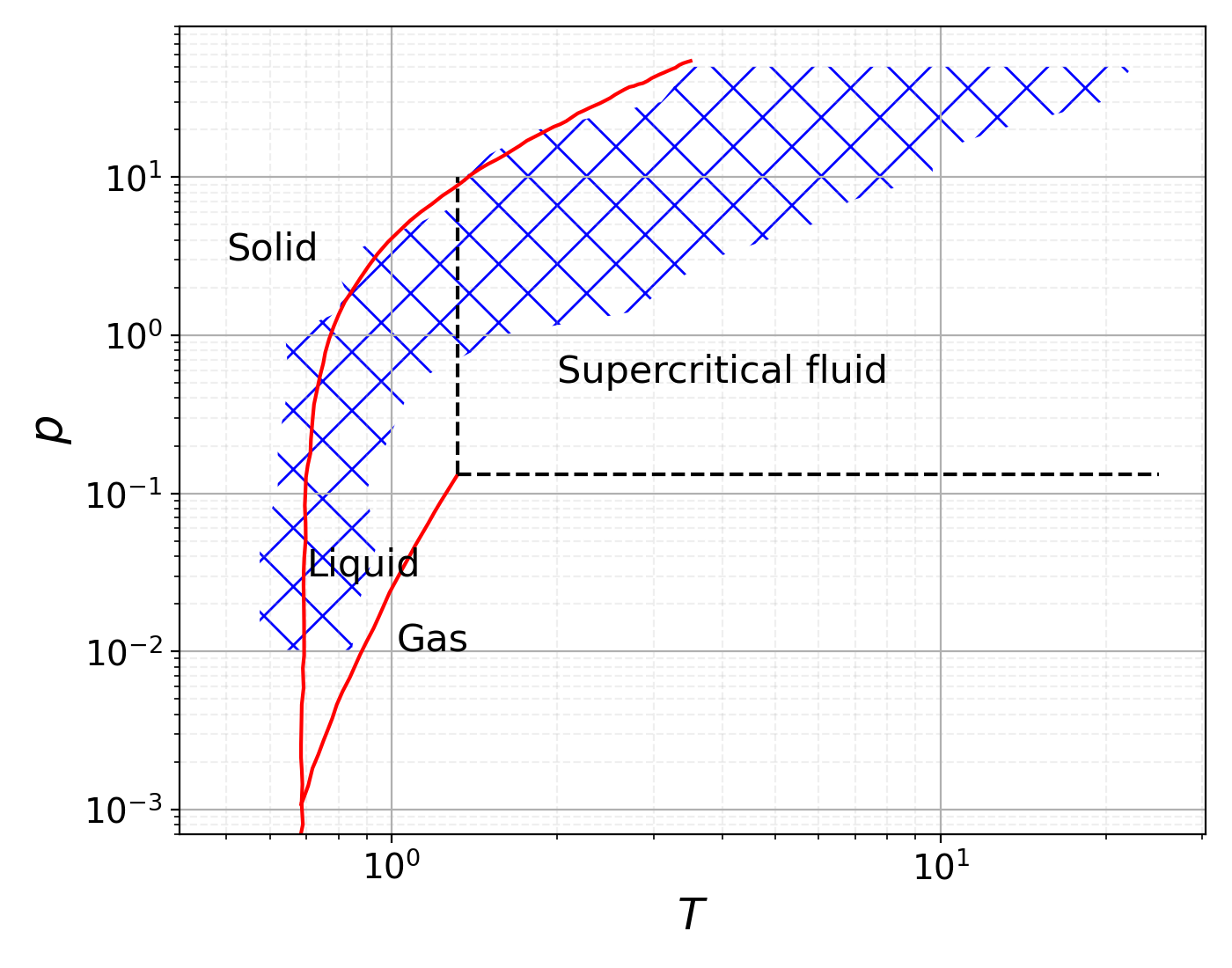}
    \caption{Pressure-temperature phase diagram of the Lennard-Jones fluid (coexistence lines are taken from~\cite{agrawal1995thermodynamic}). The blue crossed area represents the region over which viscosity is well described by the Arrhenius equation.}
    \label{fig:phasediagram_vs_arrhenius}
\end{figure}

To further characterize the range of validity of the Arrhenius model, we represent the region of the pressure-temperature phase diagram over which the Arrhenius equation holds in Fig.~\ref{fig:phasediagram_vs_arrhenius} (we considered a fit valid as long as it did not deviate from the data by more than the error bars). This region covers mainly the liquid phase, and also part of the supercritical fluid phase.

\begin{figure}
    \centering
    \includegraphics[width=0.9\linewidth]{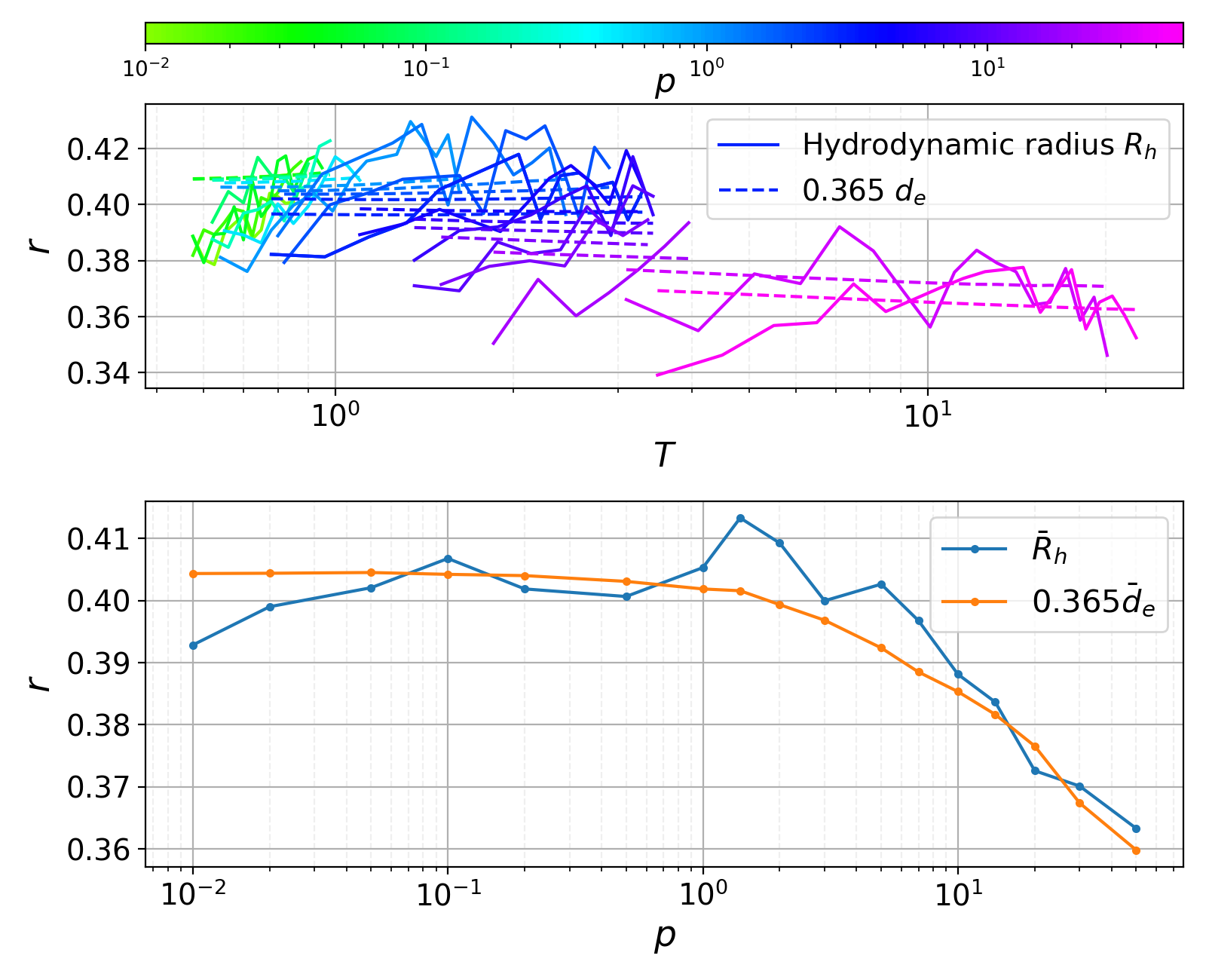}
    \caption{Testing the validity of the Stokes-Einstein law. (Top) Temperature dependence of the hydrodynamic radius $R_h$ and equilibrium distance $d_e$ at different pressures. (Bottom) Pressure dependence of the temperature-averaged hydrodynamic radius $\bar{R}_h$ and the equilibrium distance $\bar{d}_e$.}
    \label{fig:rh_vs_rc}
\end{figure}

To investigate the microscopic origin of the Arrhenian behavior of viscosity, we make a detour through atomic diffusion. This step is motivated by the idea that it is easier to build a physical picture of mass transport than one of momentum transport, and also by the possible relation between both types of transport, through the Stokes-Einstein (SE) law \cite{einstein1905molekularkinetischen,CappelezzoCapellariPezzinCoelho2007,Weiss2018,OhtoriUchiyamaIshii2018}. Indeed, according to this law, the diffusion coefficient is inversely proportional to the viscosity:
\begin{equation}
D=\frac{T}{6 \pi R_h \eta} ,
\label{eq:stokes-einstein}
\end{equation}
where $R_h$ is an effective hydrodynamic radius \cite{einstein1905molekularkinetischen,CappelezzoCapellariPezzinCoelho2007,Weiss2018,OhtoriUchiyamaIshii2018}.
The hydrodynamic radius is traditionally defined by mapping the viscous drag force $F$ on a particle moving with a velocity $U$ to the continuum calculation of Stokes for a no-slip sphere: $F = 6\pi\eta R_h U$. Note that for a single atom, one could also map the drag to that of a slipping sphere, $F = 4\pi\eta R_h^\text{slip} U$. This, however, would only change the hydrodynamic radius by a constant prefactor: $R_h^\text{slip} = \frac{3}{2} R_h$. 
In practice, we computed $R_h$ by combining the simulation data for viscosity and diffusion: $R_h = T/(6\pi\eta D)$.
For the SE law to hold, $R_h$ 
must be independent of the temperature. We verify this in Fig.~\ref{fig:rh_vs_rc}(top), where we plot the temperature dependence of the hydrodynamic radius for each of the pressures studied in Fig.~\ref{fig:visc_temp_pres}. We observe that while $R_h$ systematically decreases with increasing pressure, it varies only slightly (by less than $10\,\%$) with temperature at constant pressure, which validates the SE law. Interestingly, the hydrodynamic radius $R_h$ appears to be proportional to the equilibrium distance $d_e$ between atoms, calculated as the position of the first peak of the radial distribution function (RDF) \cite{sm}. This is illustrated in Fig.~\ref{fig:rh_vs_rc}(bottom), where the pressure dependence of the temperature-averaged hydrodynamic radius $\bar{R}_h$ is fitted by multiplying the temperature-averaged equilibrium distance $\bar{d}_e$ by a factor $0.31$.
As a side note, the slip hydrodynamic radius is then given by $\bar{R}_h^\text{slip} = \frac{3}{2} \bar{R}_h = 0.548 \bar{d}_e$, hence, it is close to half the ``static'' particle diameter, defined from the equilibrium interparticle distance.

\begin{figure}
    \centering
    \includegraphics[width=\linewidth]{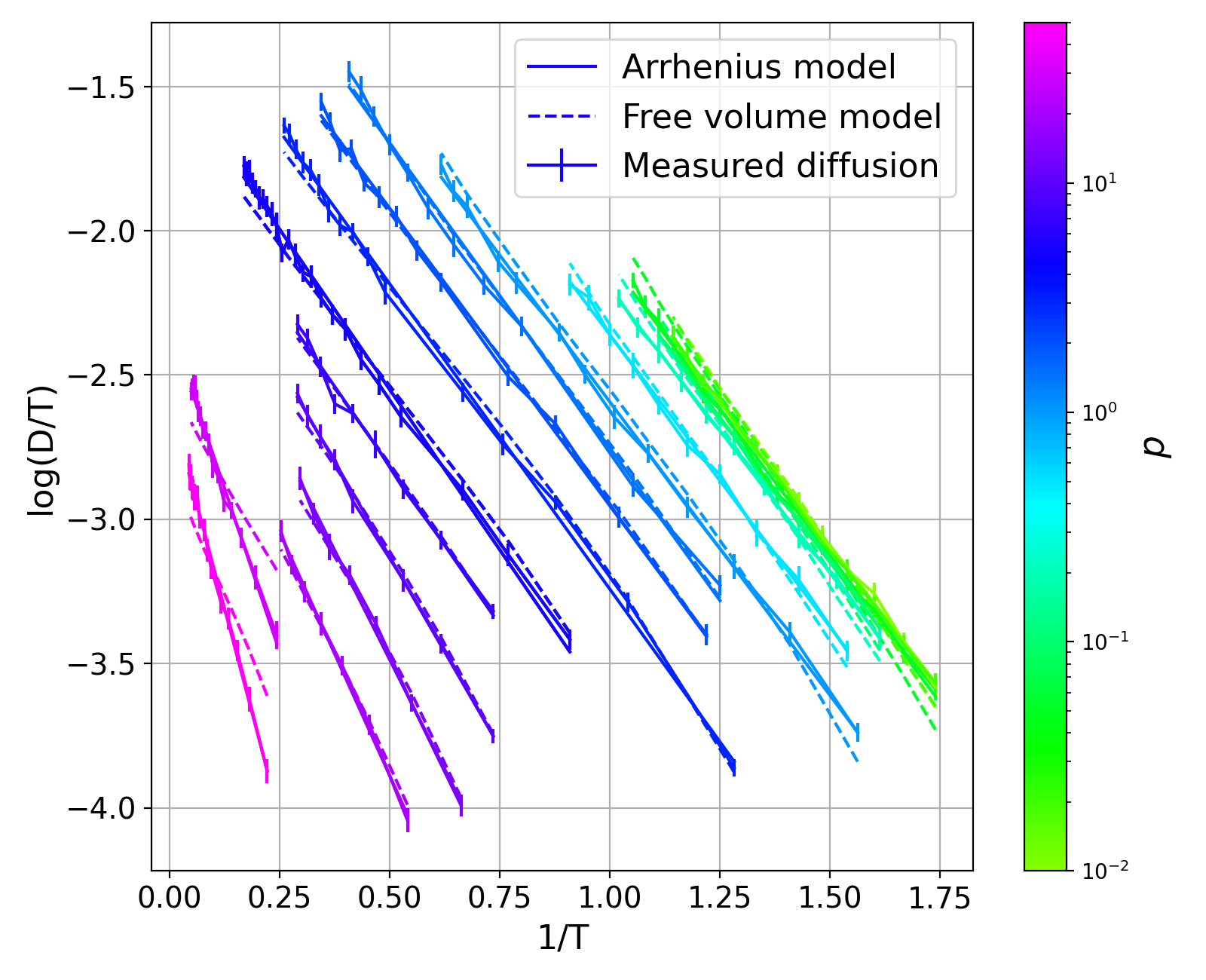}
    \caption{Comparison between the numerically measured diffusion coefficient, the Arrhenius-like model and the free volume model.}
    \label{fig:diffusion_arrhenius_freevolume}
\end{figure}

Having established the validity of the SE law, we can now proceed with the modeling of the diffusion phenomenon. 
Combining the Arrhenius law describing the viscosity, Eq.~\eqref{eq:arrh_visc}, with the SE law, Eq.~\eqref{eq:stokes-einstein}, one obtains that the diffusion coefficient is described by a pseudo-Arrhenius law:
\begin{equation}
    D=D'_0 T e^{\frac{-Q}{T}}, 
    \label{eq:arrh-diffusion}
\end{equation}
where $D'_0 = 1/(6\pi R_h \eta_0)$. 
As shown in Fig.~\ref{fig:diffusion_arrhenius_freevolume}, Eq.~\eqref{eq:arrh-diffusion} reproduces the simulation results accurately. 
Due to the observed Arrhenian behavior of viscosity and diffusion, it is tempting to turn to microscopic theories that directly encode the Arrhenius equation. The reaction rate theory or so-called \textit{Eyring}'s model, in particular, describes diffusion as a succession of local atomic rearrangements, each of which consists in the crossing of a free energy barrier \cite{Eyring1936a}. The rates at which these crossings occur are controlled by their free energy barrier through the Eyring equation. This equation is a microscopic counterpart of the Arrhenius law, and then, it is natural that an Arrhenius behavior emerges at the continuous scale from such a process. Despite this good agreement with observations, this approach suffers from various shortcomings. First, identifying local rearrangements in simple liquids is difficult as they lack any energetic bonds that would survive thermal agitation long enough so that they can be spotted. Another closely related issue lies in the fact that the activation energies obtained from the slopes of the curves in Fig.~\ref{fig:visc_temp_pres} (values reported in the SM \cite{sm}) are comparable to or even lower than the thermal energy in the range of temperature explored.
This is incompatible with the reaction rate theory, which is built on the prerequisite that the system has time to explore its local environment before crossing a barrier to a neighboring local minimum in the free energy landscape \cite{HanggiTalknerBorkovec1990}.

The inadequacy of energy-barrier-crossing models led us to investigate mass transport through the free volume model. According to this model, 
in a system of particles maintained at fixed density and under thermal agitation, collisions between particles constantly redistribute free volumes, which sometimes create voids above  a critical volume necessary 
for a particle to diffuse in. The free volume model is built upon two main parameters: the critical volume $v_c$, which is of the order of the size of the cage formed around a particle by its closest neighbors \cite{Hogenboom1967}, and the mean free volume $v_f$, which is equal to the average volume per particle, $1/\rho$, minus the average volume per particle in a state where the system's dynamics is frozen, typically the glass state. This latter average volume is called the intrinsic particle volume $v_0$ \cite{Cohen1959,Turnbull1961,Turnbull1970}. Within the free volume model \cite{Cohen1959,Turnbull1961,Turnbull1970,Falk2020}, the diffusion coefficient writes:
\begin{equation}
D=D_0\,e^{\frac{-v_c}{v_f}},
\label{eq:diffusion_freevolume}
\end{equation}
where $D_0=g v_{th}L$, with $v_{th}$ the thermal velocity of the particles, $L$ the mean free path, and $g$ an empirical geometric factor. 
The factor $g$ is often fixed at $\frac{1}{6}$ \cite{Cohen1959,Turnbull1961,Hogenboom1967,Naghizadeh1964}, but it turns in our simulation to be closer to $\frac{1}{3}$, which we will select in the following. 
One of the main critics of the free volume model lies in the empirical definition of the parameters it builds upon. However, it was recently shown that these parameters can be evaluated directly from microscopic properties in molecular, alkane-based, fluids \cite{Falk2020}. Along the same lines, we try in the following to define the different parameters entering the free volume diffusion coefficient equation, Eq.~\eqref{eq:diffusion_freevolume}, from microscopic measurements and evaluate the extent to which these nonempirical parameters can describe diffusion.

The thermal velocity $v_{th}$ is by definition 
$v_{th}=\sqrt{3T}$. The mean free path $L$ is extracted from the mean square displacement data as the distance travelled by the particle when leaving the ballistic phase (see the SM for details \cite{sm}, and results in Fig.~S5). 
$L$ is shown to be well described by the single cubic cell model \cite{Takagi1980}: $L=2\left(\rho^{-1/3} - a \right)$, 
with $\rho^{-1/3}$ the average distance between the particles and $a$ the hard-core diameter of the particles, which we identified with the contact distance $d_{c}$ measured from the RDF \cite{sm}; consequently, in the following, we will use the single cubic cell model to compute $L$.

We then estimate the mean free volume $v_f$, which is the difference between the average volume per particle $1/\rho$, set in the simulation, and the intrinsic particle volume, $v_0$; the latter is  the average volume per particle at the same pressure and temperature, but \emph{in the glass state}: $v_0 = 1/\rho_\text{glass}$. 
At low temperatures, we could obtain the glass state by rapidly quenching the liquid at constant pressure. 
We then found that $v_0$ varies as: $v_0(T;p)=v_{00}(p) \left\{1+\alpha(p)\, T \right\}$, with $v_{00}(p)$ the intrinsic particle volume at zero temperature and $\alpha$ the thermal expansion coefficient. We used this linear dependency extracted at low temperatures to extrapolate $v_0$ at any temperature 
(see details in the SM \cite{sm}). 
Finally, the critical volume, $v_c$, is fitted from our results, and we will discuss its value in the following. 

We can now systematically compare the diffusion coefficients computed from MD simulations with their estimation based on the free volume model, Eq.~\eqref{eq:diffusion_freevolume}. The results, reported in Fig.~\ref{fig:diffusion_arrhenius_freevolume}, show that the free volume model describes diffusion coefficients with high precision at low and intermediate pressures, although it becomes less accurate at the highest pressures ($p=30, 50$).
The critical volume $v_c$ is found to be on the order of $0.1$ at low pressure, and to decrease with increasing pressure  \cite{sm}. This is in contrast with usual values reported in the literature \cite{Falk2020, Hogenboom1967, Cohen1959}, which are generally closer to the intrinsic particle volume $v_0$. 
In particular, the procedure proposed by \citet{Falk2020} to estimate $v_c$ from simple structural parameters in the context of molecular liquids largely overestimates the values of $v_c$ we obtained from the fits, and does not reproduce the decrease of $v_c$ with pressure that we observed. 
As a consequence of the small values of $v_c$ for the LJ fluid, the exponential term in the free volume model, $\exp(-v_c/v_f)$, is close to one, i.e., collisions in the fluid very often succeed in causing diffusion. Note that, in this case, there is no fundamental issue with the exponential term being close to one, and no incompatibility with the underlying theory, in contrast with the \textit{Eyring}'s model discussed above.

We are now equipped with two models to describe the diffusion coefficient and the viscosity -- related through the SE equation -- 
with acceptable precision. Interestingly, these two models are each constructed on a very different physical description of mass transport. The fact that it is possible to recover an Arrhenian behavior with the free volume model further confirms that the physical picture at the origin of this behavior is not necessarily one based on energy-barrier-crossing events. 
The coincidental nature of the match between the predictions of the two models is also emphasized in Fig.~S10 of the SM \cite{sm}, which shows that there is no relation between the exponential terms of the models, neither in terms of amplitude nor in terms of variation.

\paragraph*{Conclusion--}
We have run MD simulations to compute the viscosity and the diffusion coefficient of a simple LJ fluid for different temperatures and pressures. We observed that the viscosity is accurately described by the Arrhenius law in a wide range of temperatures and pressures in the liquid and supercritical fluid phases. To understand the microscopic origin of this Arrhenian behavior we made a detour through the study of the diffusion coefficient. We verified that the viscosity and diffusion coefficient of the LJ fluid are correlated through the SE law, with a temperature-independent hydrodynamic radius $R_h$ that is linked to the interparticle equilibrium distance $d_e$. Even though the Arrhenius equation reproduces the computed viscosity and diffusion data, it cannot be derived from reaction rate theories as these are not suitable to describe the liquid state, in which activation energies are on the order of or larger than thermal energies. For this reason, we have considered the free volume model, which also described well the evolution of the diffusion coefficient with temperature and pressure. 
We have shown that the parameters of the free volume model can be measured independently, more specifically, the mean free path, 
and the average free volume -- calculated from the fluid density and the density of the glass at the same pressure and temperature. This led us to have only the critical volume as a fitting parameter. 
Overall, the free volume model appears more justified from a microscopic point of view because one can estimate its parameters from microscopic properties. 

In future work, it would be interesting to explore how the above discussion on the microscopic origin of the Arrhenius law observed for a LJ liquid extends to more complex situations, and in particular to real liquids of interest such as water. Additionally, exploring the connection between the rather simple free volume model, and alternative microscopic descriptions such as excess entropy scaling \cite{Rosenfeld1977,Rosenfeld1999,Galliero2011,Bell2019,Viet2022} 
could provide additional insights on the emergence of a simple Arrhenius behaviour without the need for activated processes. 
Finally, rationalizing the deviations from the Arrhenius behavior typical of fragile supercooled liquids \cite{Tarjus2004,Hentschel2012,parmar2020stable}, where thermal energy can become lower than activation energies, could require going one step further and combining both free volume and activation models \cite{MacedoLitovitz1965}.

\begin{acknowledgments}
The authors thank Cecilia Herrero and Kerstin Falk for fruitful discussions. 
We are also grateful for HPC resources
from GENCI/TGCC (grant A0090810637), and from the PSMN mesocenter in Lyon. 
LJ is supported by the Institut Universitaire de France. 
\end{acknowledgments}


\input{main.bbl}

\end{document}

%% file: main.bbl
%

%% file: main.bbl
\begin{thebibliography}{40}%
\makeatletter
\providecommand \@ifxundefined [1]{%
 \@ifx{#1\undefined}
}%
\providecommand \@ifnum [1]{%
 \ifnum #1\expandafter \@firstoftwo
 \else \expandafter \@secondoftwo
 \fi
}%
\providecommand \@ifx [1]{%
 \ifx #1\expandafter \@firstoftwo
 \else \expandafter \@secondoftwo
 \fi
}%
\providecommand \natexlab [1]{#1}%
\providecommand \enquote  [1]{``#1''}%
\providecommand \bibnamefont  [1]{#1}%
\providecommand \bibfnamefont [1]{#1}%
\providecommand \citenamefont [1]{#1}%
\providecommand \href@noop [0]{\@secondoftwo}%
\providecommand \href [0]{\begingroup \@sanitize@url \@href}%
\providecommand \@href[1]{\@@startlink{#1}\@@href}%
\providecommand \@@href[1]{\endgroup#1\@@endlink}%
\providecommand \@sanitize@url [0]{\catcode `\\12\catcode `\$12\catcode
  `\&12\catcode `\#12\catcode `\^12\catcode `\_12\catcode `\%12\relax}%
\providecommand \@@startlink[1]{}%
\providecommand \@@endlink[0]{}%
\providecommand \url  [0]{\begingroup\@sanitize@url \@url }%
\providecommand \@url [1]{\endgroup\@href {#1}{\urlprefix }}%
\providecommand \urlprefix  [0]{URL }%
\providecommand \Eprint [0]{\href }%
\providecommand \doibase [0]{https://doi.org/}%
\providecommand \selectlanguage [0]{\@gobble}%
\providecommand \bibinfo  [0]{\@secondoftwo}%
\providecommand \bibfield  [0]{\@secondoftwo}%
\providecommand \translation [1]{[#1]}%
\providecommand \BibitemOpen [0]{}%
\providecommand \bibitemStop [0]{}%
\providecommand \bibitemNoStop [0]{.\EOS\space}%
\providecommand \EOS [0]{\spacefactor3000\relax}%
\providecommand \BibitemShut  [1]{\csname bibitem#1\endcsname}%
\let\auto@bib@innerbib\@empty
\bibitem [{\citenamefont {Henne}\ \emph {et~al.}(2004)\citenamefont {Henne},
  \citenamefont {Breyer}, \citenamefont {Niedermeier},\ and\ \citenamefont
  {Ermanni}}]{Henne2004}%
  \BibitemOpen
  \bibfield  {author} {\bibinfo {author} {\bibfnamefont {M.}~\bibnamefont
  {Henne}}, \bibinfo {author} {\bibfnamefont {C.}~\bibnamefont {Breyer}},
  \bibinfo {author} {\bibfnamefont {M.}~\bibnamefont {Niedermeier}},\ and\
  \bibinfo {author} {\bibfnamefont {P.}~\bibnamefont {Ermanni}},\ }\bibfield
  {title} {\bibinfo {title} {{A new kinetic and viscosity model for liquid
  composite molding simulations in an industrial environment}},\ }\href
  {https://doi.org/10.1002/pc.20020} {\bibfield  {journal} {\bibinfo  {journal}
  {Polym. Compos.}\ }\textbf {\bibinfo {volume} {25}},\ \bibinfo {pages} {255}
  (\bibinfo {year} {2004})}\BibitemShut {NoStop}%
\bibitem [{\citenamefont {Quinchia}\ \emph {et~al.}(2010)\citenamefont
  {Quinchia}, \citenamefont {Delgado}, \citenamefont {Valencia}, \citenamefont
  {Franco},\ and\ \citenamefont {Gallegos}}]{Quinchia2010}%
  \BibitemOpen
  \bibfield  {author} {\bibinfo {author} {\bibfnamefont {L.~A.}\ \bibnamefont
  {Quinchia}}, \bibinfo {author} {\bibfnamefont {M.~A.}\ \bibnamefont
  {Delgado}}, \bibinfo {author} {\bibfnamefont {C.}~\bibnamefont {Valencia}},
  \bibinfo {author} {\bibfnamefont {J.~M.}\ \bibnamefont {Franco}},\ and\
  \bibinfo {author} {\bibfnamefont {C.}~\bibnamefont {Gallegos}},\ }\bibfield
  {title} {\bibinfo {title} {{Viscosity modification of different vegetable
  oils with EVA copolymer for lubricant applications}},\ }\href
  {https://doi.org/10.1016/j.indcrop.2010.07.011} {\bibfield  {journal}
  {\bibinfo  {journal} {Ind. Crops Prod.}\ }\textbf {\bibinfo {volume} {32}},\
  \bibinfo {pages} {607} (\bibinfo {year} {2010})}\BibitemShut {NoStop}%
\bibitem [{\citenamefont {{M. Takechi, C. Uno}}(1994)}]{M.TakechiC.Uno1994}%
  \BibitemOpen
  \bibfield  {author} {\bibinfo {author} {\bibfnamefont {Y.~T.}\ \bibnamefont
  {{M. Takechi, C. Uno}}},\ }\bibfield  {title} {\bibinfo {title}
  {{NII-Electronic Library Service}},\ }\href
  {https://www.jstage.jst.go.jp/article/bpb1993/17/11/17{\_}11{\_}1460/{\_}pdf/-char/ja}
  {\bibfield  {journal} {\bibinfo  {journal} {Chem. Pharm. Bull.}\ }\textbf
  {\bibinfo {volume} {17}},\ \bibinfo {pages} {1460} (\bibinfo {year}
  {1994})}\BibitemShut {NoStop}%
\bibitem [{\citenamefont {Andrade}(1930)}]{andrade1930viscosity}%
  \BibitemOpen
  \bibfield  {author} {\bibinfo {author} {\bibfnamefont {E.~d.~C.}\
  \bibnamefont {Andrade}},\ }\bibfield  {title} {\bibinfo {title} {The
  viscosity of liquids},\ }\href@noop {} {\bibfield  {journal} {\bibinfo
  {journal} {Nature}\ }\textbf {\bibinfo {volume} {125}},\ \bibinfo {pages}
  {309} (\bibinfo {year} {1930})}\BibitemShut {NoStop}%
\bibitem [{\citenamefont {Davies}\ and\ \citenamefont
  {Matheson}(1967)}]{DaviesMatheson1967}%
  \BibitemOpen
  \bibfield  {author} {\bibinfo {author} {\bibfnamefont {D.~B.}\ \bibnamefont
  {Davies}}\ and\ \bibinfo {author} {\bibfnamefont {A.~J.}\ \bibnamefont
  {Matheson}},\ }\bibfield  {title} {\bibinfo {title} {Viscosity of liquids
  containing spherical molecules or ions},\ }\href
  {https://doi.org/10.1039/tf9676300596} {\bibfield  {journal} {\bibinfo
  {journal} {Transactions of the Faraday Society}\ }\textbf {\bibinfo {volume}
  {63}},\ \bibinfo {pages} {596} (\bibinfo {year} {1967})}\BibitemShut
  {NoStop}%
\bibitem [{\citenamefont {Poirier}(1988)}]{Poirier1988}%
  \BibitemOpen
  \bibfield  {author} {\bibinfo {author} {\bibfnamefont {J.~P.}\ \bibnamefont
  {Poirier}},\ }\bibfield  {title} {\bibinfo {title} {Transport properties of
  liquid metals and viscosity of the earth's core},\ }\href
  {https://doi.org/10.1111/j.1365-246x.1988.tb01124.x} {\bibfield  {journal}
  {\bibinfo  {journal} {Geophysical Journal International}\ }\textbf {\bibinfo
  {volume} {92}},\ \bibinfo {pages} {99} (\bibinfo {year} {1988})}\BibitemShut
  {NoStop}%
\bibitem [{\citenamefont {Messa{\^{a}}di}\ \emph {et~al.}(2015)\citenamefont
  {Messa{\^{a}}di}, \citenamefont {Dhouibi}, \citenamefont {Hamda},
  \citenamefont {Belgacem}, \citenamefont {Adbelkader}, \citenamefont
  {Ouerfelli},\ and\ \citenamefont
  {Hamzaoui}}]{MessaadiDhouibiHamdaBelgacemAdbelkaderOuerfelliHamzaoui2015}%
  \BibitemOpen
  \bibfield  {author} {\bibinfo {author} {\bibfnamefont {A.}~\bibnamefont
  {Messa{\^{a}}di}}, \bibinfo {author} {\bibfnamefont {N.}~\bibnamefont
  {Dhouibi}}, \bibinfo {author} {\bibfnamefont {H.}~\bibnamefont {Hamda}},
  \bibinfo {author} {\bibfnamefont {F.~B.~M.}\ \bibnamefont {Belgacem}},
  \bibinfo {author} {\bibfnamefont {Y.~H.}\ \bibnamefont {Adbelkader}},
  \bibinfo {author} {\bibfnamefont {N.}~\bibnamefont {Ouerfelli}},\ and\
  \bibinfo {author} {\bibfnamefont {A.~H.}\ \bibnamefont {Hamzaoui}},\
  }\bibfield  {title} {\bibinfo {title} {A new equation relating the viscosity
  arrhenius temperature and the activation energy for some newtonian classical
  solvents},\ }\href {https://doi.org/10.1155/2015/163262} {\bibfield
  {journal} {\bibinfo  {journal} {Journal of Chemistry}\ }\textbf {\bibinfo
  {volume} {2015}},\ \bibinfo {pages} {1} (\bibinfo {year} {2015})}\BibitemShut
  {NoStop}%
\bibitem [{\citenamefont {Eyring}(1936)}]{Eyring1936a}%
  \BibitemOpen
  \bibfield  {author} {\bibinfo {author} {\bibfnamefont {H.}~\bibnamefont
  {Eyring}},\ }\bibfield  {title} {\bibinfo {title} {{Viscosity, plasticity,
  and diffusion as examples of absolute reaction rates}},\ }\href
  {https://doi.org/10.1063/1.1749836} {\bibfield  {journal} {\bibinfo
  {journal} {J. Chem. Phys.}\ }\textbf {\bibinfo {volume} {4}},\ \bibinfo
  {pages} {283} (\bibinfo {year} {1936})}\BibitemShut {NoStop}%
\bibitem [{\citenamefont {Mac{\'{\i}}as-Salinas}\ \emph
  {et~al.}(2008)\citenamefont {Mac{\'{\i}}as-Salinas}, \citenamefont
  {Dur{\'{a}}n-Valencia}, \citenamefont {L{\'{o}}pez-Ram{\'{\i}}rez},\ and\
  \citenamefont {Bouchot}}]{MaciasSalinasDuranValenciaLopezRamirezBouchot2008}%
  \BibitemOpen
  \bibfield  {author} {\bibinfo {author} {\bibfnamefont {R.}~\bibnamefont
  {Mac{\'{\i}}as-Salinas}}, \bibinfo {author} {\bibfnamefont {C.}~\bibnamefont
  {Dur{\'{a}}n-Valencia}}, \bibinfo {author} {\bibfnamefont {S.}~\bibnamefont
  {L{\'{o}}pez-Ram{\'{\i}}rez}},\ and\ \bibinfo {author} {\bibfnamefont
  {C.}~\bibnamefont {Bouchot}},\ }\bibfield  {title} {\bibinfo {title}
  {Eyring-theory-based model to estimate crude oil viscosity at reservoir
  conditions},\ }\href {https://doi.org/10.1021/ef8003015} {\bibfield
  {journal} {\bibinfo  {journal} {Energy {\&} Fuels}\ }\textbf {\bibinfo
  {volume} {23}},\ \bibinfo {pages} {464} (\bibinfo {year} {2008})}\BibitemShut
  {NoStop}%
\bibitem [{\citenamefont {Heyes}\ \emph {et~al.}(2018)\citenamefont {Heyes},
  \citenamefont {Dini},\ and\ \citenamefont {Smith}}]{HeyesDiniSmith2018}%
  \BibitemOpen
  \bibfield  {author} {\bibinfo {author} {\bibfnamefont {D.~M.}\ \bibnamefont
  {Heyes}}, \bibinfo {author} {\bibfnamefont {D.}~\bibnamefont {Dini}},\ and\
  \bibinfo {author} {\bibfnamefont {E.~R.}\ \bibnamefont {Smith}},\ }\bibfield
  {title} {\bibinfo {title} {Incremental viscosity by non-equilibrium molecular
  dynamics and the {E}yring model},\ }\href {https://doi.org/10.1063/1.5027681}
  {\bibfield  {journal} {\bibinfo  {journal} {The Journal of Chemical Physics}\
  }\textbf {\bibinfo {volume} {148}},\ \bibinfo {pages} {194506} (\bibinfo
  {year} {2018})}\BibitemShut {NoStop}%
\bibitem [{\citenamefont {Cohen}\ and\ \citenamefont
  {Turnbull}(1959)}]{Cohen1959}%
  \BibitemOpen
  \bibfield  {author} {\bibinfo {author} {\bibfnamefont {M.~H.}\ \bibnamefont
  {Cohen}}\ and\ \bibinfo {author} {\bibfnamefont {D.}~\bibnamefont
  {Turnbull}},\ }\bibfield  {title} {\bibinfo {title} {{Molecular transport in
  liquids and glasses}},\ }\href {https://doi.org/10.1063/1.1730566} {\bibfield
   {journal} {\bibinfo  {journal} {J. Chem. Phys.}\ }\textbf {\bibinfo {volume}
  {31}},\ \bibinfo {pages} {1164} (\bibinfo {year} {1959})}\BibitemShut
  {NoStop}%
\bibitem [{\citenamefont {Turnbull}\ and\ \citenamefont
  {Cohen}(1961)}]{Turnbull1961}%
  \BibitemOpen
  \bibfield  {author} {\bibinfo {author} {\bibfnamefont {D.}~\bibnamefont
  {Turnbull}}\ and\ \bibinfo {author} {\bibfnamefont {M.~H.}\ \bibnamefont
  {Cohen}},\ }\bibfield  {title} {\bibinfo {title} {{Free-volume model of the
  amorphous phase: Glass transition}},\ }\href
  {https://doi.org/10.1063/1.1731549} {\bibfield  {journal} {\bibinfo
  {journal} {J. Chem. Phys.}\ }\textbf {\bibinfo {volume} {34}},\ \bibinfo
  {pages} {120} (\bibinfo {year} {1961})}\BibitemShut {NoStop}%
\bibitem [{\citenamefont {Turnbull}\ and\ \citenamefont
  {Cohen}(1970)}]{Turnbull1970}%
  \BibitemOpen
  \bibfield  {author} {\bibinfo {author} {\bibfnamefont {D.}~\bibnamefont
  {Turnbull}}\ and\ \bibinfo {author} {\bibfnamefont {M.~H.}\ \bibnamefont
  {Cohen}},\ }\bibfield  {title} {\bibinfo {title} {{On the free-volume model
  of the liquid-glass transition}},\ }\href {https://doi.org/10.1063/1.1673434}
  {\bibfield  {journal} {\bibinfo  {journal} {J. Chem. Phys.}\ }\textbf
  {\bibinfo {volume} {52}},\ \bibinfo {pages} {3038} (\bibinfo {year}
  {1970})}\BibitemShut {NoStop}%
\bibitem [{\citenamefont {Falk}\ \emph {et~al.}(2020)\citenamefont {Falk},
  \citenamefont {Savio},\ and\ \citenamefont {Moseler}}]{Falk2020}%
  \BibitemOpen
  \bibfield  {author} {\bibinfo {author} {\bibfnamefont {K.}~\bibnamefont
  {Falk}}, \bibinfo {author} {\bibfnamefont {D.}~\bibnamefont {Savio}},\ and\
  \bibinfo {author} {\bibfnamefont {M.}~\bibnamefont {Moseler}},\ }\bibfield
  {title} {\bibinfo {title} {{Nonempirical Free Volume Viscosity Model for
  Alkane Lubricants under Severe Pressures}},\ }\href
  {https://doi.org/10.1103/PhysRevLett.124.105501} {\bibfield  {journal}
  {\bibinfo  {journal} {Phys. Rev. Lett.}\ }\textbf {\bibinfo {volume} {124}},\
  \bibinfo {pages} {105501} (\bibinfo {year} {2020})},\ \Eprint
  {https://arxiv.org/abs/1905.06130} {arXiv:1905.06130} \BibitemShut {NoStop}%
\bibitem [{\citenamefont {Dyre}(2014)}]{Dyre2014}%
  \BibitemOpen
  \bibfield  {author} {\bibinfo {author} {\bibfnamefont {J.~C.}\ \bibnamefont
  {Dyre}},\ }\bibfield  {title} {\bibinfo {title} {Hidden scale invariance in
  condensed matter},\ }\href {https://doi.org/10.1021/jp501852b} {\bibfield
  {journal} {\bibinfo  {journal} {The Journal of Physical Chemistry B}\
  }\textbf {\bibinfo {volume} {118}},\ \bibinfo {pages} {10007} (\bibinfo
  {year} {2014})}\BibitemShut {NoStop}%
\bibitem [{\citenamefont {Liu}\ \emph {et~al.}(1998)\citenamefont {Liu},
  \citenamefont {Silva},\ and\ \citenamefont {Macedo}}]{Liu1998}%
  \BibitemOpen
  \bibfield  {author} {\bibinfo {author} {\bibfnamefont {H.}~\bibnamefont
  {Liu}}, \bibinfo {author} {\bibfnamefont {C.~M.}\ \bibnamefont {Silva}},\
  and\ \bibinfo {author} {\bibfnamefont {E.~A.}\ \bibnamefont {Macedo}},\
  }\bibfield  {title} {\bibinfo {title} {{Unified approach to the
  self-diffusion coefficients of dense fluids over wide ranges of temperature
  and pressure - Hard-sphere, square-well, Lennard-Jones and real
  substances}},\ }\href {https://doi.org/10.1016/S0009-2509(98)00036-0}
  {\bibfield  {journal} {\bibinfo  {journal} {Chem. Eng. Sci.}\ }\textbf
  {\bibinfo {volume} {53}},\ \bibinfo {pages} {2403} (\bibinfo {year}
  {1998})}\BibitemShut {NoStop}%
\bibitem [{\citenamefont {Galli{\'{e}}ro}\ \emph {et~al.}(2005)\citenamefont
  {Galli{\'{e}}ro}, \citenamefont {Boned},\ and\ \citenamefont
  {Baylaucq}}]{Galliero2005}%
  \BibitemOpen
  \bibfield  {author} {\bibinfo {author} {\bibfnamefont {G.}~\bibnamefont
  {Galli{\'{e}}ro}}, \bibinfo {author} {\bibfnamefont {C.}~\bibnamefont
  {Boned}},\ and\ \bibinfo {author} {\bibfnamefont {A.}~\bibnamefont
  {Baylaucq}},\ }\bibfield  {title} {\bibinfo {title} {{Molecular dynamics
  study of the Lennard-Jones fluid viscosity: Application to real fluids}},\
  }\href {https://doi.org/10.1021/ie050154t} {\bibfield  {journal} {\bibinfo
  {journal} {Ind. Eng. Chem. Res.}\ }\textbf {\bibinfo {volume} {44}},\
  \bibinfo {pages} {6963} (\bibinfo {year} {2005})}\BibitemShut {NoStop}%
\bibitem [{\citenamefont {Thompson}\ \emph {et~al.}(2022)\citenamefont
  {Thompson}, \citenamefont {Aktulga}, \citenamefont {Berger}, \citenamefont
  {Bolintineanu}, \citenamefont {Brown}, \citenamefont {Crozier}, \citenamefont
  {{in 't Veld}}, \citenamefont {Kohlmeyer}, \citenamefont {Moore},
  \citenamefont {Nguyen}, \citenamefont {Shan}, \citenamefont {Stevens},
  \citenamefont {Tranchida}, \citenamefont {Trott},\ and\ \citenamefont
  {Plimpton}}]{LAMMPS2022}%
  \BibitemOpen
  \bibfield  {author} {\bibinfo {author} {\bibfnamefont {A.~P.}\ \bibnamefont
  {Thompson}}, \bibinfo {author} {\bibfnamefont {H.~M.}\ \bibnamefont
  {Aktulga}}, \bibinfo {author} {\bibfnamefont {R.}~\bibnamefont {Berger}},
  \bibinfo {author} {\bibfnamefont {D.~S.}\ \bibnamefont {Bolintineanu}},
  \bibinfo {author} {\bibfnamefont {W.~M.}\ \bibnamefont {Brown}}, \bibinfo
  {author} {\bibfnamefont {P.~S.}\ \bibnamefont {Crozier}}, \bibinfo {author}
  {\bibfnamefont {P.~J.}\ \bibnamefont {{in 't Veld}}}, \bibinfo {author}
  {\bibfnamefont {A.}~\bibnamefont {Kohlmeyer}}, \bibinfo {author}
  {\bibfnamefont {S.~G.}\ \bibnamefont {Moore}}, \bibinfo {author}
  {\bibfnamefont {T.~D.}\ \bibnamefont {Nguyen}}, \bibinfo {author}
  {\bibfnamefont {R.}~\bibnamefont {Shan}}, \bibinfo {author} {\bibfnamefont
  {M.~J.}\ \bibnamefont {Stevens}}, \bibinfo {author} {\bibfnamefont
  {J.}~\bibnamefont {Tranchida}}, \bibinfo {author} {\bibfnamefont
  {C.}~\bibnamefont {Trott}},\ and\ \bibinfo {author} {\bibfnamefont {S.~J.}\
  \bibnamefont {Plimpton}},\ }\bibfield  {title} {\bibinfo {title} {{LAMMPS - a
  flexible simulation tool for particle-based materials modeling at the atomic,
  meso, and continuum scales}},\ }\href
  {https://doi.org/10.1016/j.cpc.2021.108171} {\bibfield  {journal} {\bibinfo
  {journal} {Comput. Phys. Commun.}\ }\textbf {\bibinfo {volume} {271}},\
  \bibinfo {pages} {108171} (\bibinfo {year} {2022})}\BibitemShut {NoStop}%
\bibitem [{sm()}]{sm}%
  \BibitemOpen
  \href@noop {} {}\bibinfo {note} {See Supplemental Material at [URL] for
  further details.}\BibitemShut {Stop}%
\bibitem [{\citenamefont {Allen}\ and\ \citenamefont
  {Tildesley}(2017)}]{allen2017computer}%
  \BibitemOpen
  \bibfield  {author} {\bibinfo {author} {\bibfnamefont {M.~P.}\ \bibnamefont
  {Allen}}\ and\ \bibinfo {author} {\bibfnamefont {D.~J.}\ \bibnamefont
  {Tildesley}},\ }\href@noop {} {\emph {\bibinfo {title} {Computer simulation
  of liquids}}}\ (\bibinfo  {publisher} {Oxford university press},\ \bibinfo
  {year} {2017})\BibitemShut {NoStop}%
\bibitem [{\citenamefont {Heyes}\ \emph {et~al.}(2019)\citenamefont {Heyes},
  \citenamefont {Smith},\ and\ \citenamefont {Dini}}]{Heyes2019}%
  \BibitemOpen
  \bibfield  {author} {\bibinfo {author} {\bibfnamefont {D.~M.}\ \bibnamefont
  {Heyes}}, \bibinfo {author} {\bibfnamefont {E.~R.}\ \bibnamefont {Smith}},\
  and\ \bibinfo {author} {\bibfnamefont {D.}~\bibnamefont {Dini}},\ }\bibfield
  {title} {\bibinfo {title} {{Shear stress relaxation and diffusion in simple
  liquids by molecular dynamics simulations: Analytic expressions and paths to
  viscosity}},\ }\href {https://doi.org/10.1063/1.5095501} {\bibfield
  {journal} {\bibinfo  {journal} {J. Chem. Phys.}\ }\textbf {\bibinfo {volume}
  {150}},\ \bibinfo {pages} {174504} (\bibinfo {year} {2019})}\BibitemShut
  {NoStop}%
\bibitem [{\citenamefont {Carvalho-Silva}\ \emph {et~al.}(2019)\citenamefont
  {Carvalho-Silva}, \citenamefont {Coutinho},\ and\ \citenamefont
  {Aquilanti}}]{Carvalho-Silva2019}%
  \BibitemOpen
  \bibfield  {author} {\bibinfo {author} {\bibfnamefont {V.~H.}\ \bibnamefont
  {Carvalho-Silva}}, \bibinfo {author} {\bibfnamefont {N.~D.}\ \bibnamefont
  {Coutinho}},\ and\ \bibinfo {author} {\bibfnamefont {V.}~\bibnamefont
  {Aquilanti}},\ }\bibfield  {title} {\bibinfo {title} {{Temperature Dependence
  of Rate Processes Beyond Arrhenius and Eyring: Activation and
  Transitivity}},\ }\href {https://doi.org/10.3389/fchem.2019.00380} {\bibfield
   {journal} {\bibinfo  {journal} {Front. Chem.}\ }\textbf {\bibinfo {volume}
  {7}},\ \bibinfo {pages} {1} (\bibinfo {year} {2019})}\BibitemShut {NoStop}%
\bibitem [{\citenamefont {Agrawal}\ and\ \citenamefont
  {Kofke}(1995)}]{agrawal1995thermodynamic}%
  \BibitemOpen
  \bibfield  {author} {\bibinfo {author} {\bibfnamefont {R.}~\bibnamefont
  {Agrawal}}\ and\ \bibinfo {author} {\bibfnamefont {D.~A.}\ \bibnamefont
  {Kofke}},\ }\bibfield  {title} {\bibinfo {title} {Thermodynamic and
  structural properties of model systems at solid-fluid coexistence: I. fcc and
  bcc soft spheres},\ }\href@noop {} {\bibfield  {journal} {\bibinfo  {journal}
  {Molecular physics}\ }\textbf {\bibinfo {volume} {85}},\ \bibinfo {pages}
  {23} (\bibinfo {year} {1995})}\BibitemShut {NoStop}%
\bibitem [{\citenamefont {Einstein}(1905)}]{einstein1905molekularkinetischen}%
  \BibitemOpen
  \bibfield  {author} {\bibinfo {author} {\bibfnamefont {A.}~\bibnamefont
  {Einstein}},\ }\bibfield  {title} {\bibinfo {title} {{\"U}ber die von der
  molekularkinetischen theorie der w{\"a}rme geforderte bewegung von in
  ruhenden fl{\"u}ssigkeiten suspendierten teilchen},\ }\href@noop {}
  {\bibfield  {journal} {\bibinfo  {journal} {Annalen der physik}\ }\textbf
  {\bibinfo {volume} {4}} (\bibinfo {year} {1905})}\BibitemShut {NoStop}%
\bibitem [{\citenamefont {Cappelezzo}\ \emph {et~al.}(2007)\citenamefont
  {Cappelezzo}, \citenamefont {Capellari}, \citenamefont {Pezzin},\ and\
  \citenamefont {Coelho}}]{CappelezzoCapellariPezzinCoelho2007}%
  \BibitemOpen
  \bibfield  {author} {\bibinfo {author} {\bibfnamefont {M.}~\bibnamefont
  {Cappelezzo}}, \bibinfo {author} {\bibfnamefont {C.~A.}\ \bibnamefont
  {Capellari}}, \bibinfo {author} {\bibfnamefont {S.~H.}\ \bibnamefont
  {Pezzin}},\ and\ \bibinfo {author} {\bibfnamefont {L.~A.~F.}\ \bibnamefont
  {Coelho}},\ }\bibfield  {title} {\bibinfo {title} {Stokes-einstein relation
  for pure simple fluids},\ }\href {https://doi.org/10.1063/1.2738063}
  {\bibfield  {journal} {\bibinfo  {journal} {The Journal of Chemical Physics}\
  }\textbf {\bibinfo {volume} {126}},\ \bibinfo {pages} {224516} (\bibinfo
  {year} {2007})}\BibitemShut {NoStop}%
\bibitem [{\citenamefont {Weiss}\ \emph {et~al.}(2018)\citenamefont {Weiss},
  \citenamefont {Dahirel}, \citenamefont {Marry},\ and\ \citenamefont
  {Jardat}}]{Weiss2018}%
  \BibitemOpen
  \bibfield  {author} {\bibinfo {author} {\bibfnamefont {L.~B.}\ \bibnamefont
  {Weiss}}, \bibinfo {author} {\bibfnamefont {V.}~\bibnamefont {Dahirel}},
  \bibinfo {author} {\bibfnamefont {V.}~\bibnamefont {Marry}},\ and\ \bibinfo
  {author} {\bibfnamefont {M.}~\bibnamefont {Jardat}},\ }\bibfield  {title}
  {\bibinfo {title} {{Computation of the Hydrodynamic Radius of Charged
  Nanoparticles from Nonequilibrium Molecular Dynamics}},\ }\href
  {https://doi.org/10.1021/acs.jpcb.8b01153} {\bibfield  {journal} {\bibinfo
  {journal} {J. Phys. Chem. B}\ }\textbf {\bibinfo {volume} {122}},\ \bibinfo
  {pages} {5940} (\bibinfo {year} {2018})},\ \Eprint
  {https://arxiv.org/abs/1805.03478} {arXiv:1805.03478} \BibitemShut {NoStop}%
\bibitem [{\citenamefont {Ohtori}\ \emph {et~al.}(2018)\citenamefont {Ohtori},
  \citenamefont {Uchiyama},\ and\ \citenamefont
  {Ishii}}]{OhtoriUchiyamaIshii2018}%
  \BibitemOpen
  \bibfield  {author} {\bibinfo {author} {\bibfnamefont {N.}~\bibnamefont
  {Ohtori}}, \bibinfo {author} {\bibfnamefont {H.}~\bibnamefont {Uchiyama}},\
  and\ \bibinfo {author} {\bibfnamefont {Y.}~\bibnamefont {Ishii}},\ }\bibfield
   {title} {\bibinfo {title} {The stokes-einstein relation for simple fluids:
  From hard-sphere to lennard-jones via {WCA} potentials},\ }\href
  {https://doi.org/10.1063/1.5054577} {\bibfield  {journal} {\bibinfo
  {journal} {The Journal of Chemical Physics}\ }\textbf {\bibinfo {volume}
  {149}},\ \bibinfo {pages} {214501} (\bibinfo {year} {2018})}\BibitemShut
  {NoStop}%
\bibitem [{\citenamefont {H\"anggi}\ \emph {et~al.}(1990)\citenamefont
  {H\"anggi}, \citenamefont {Talkner},\ and\ \citenamefont
  {Borkovec}}]{HanggiTalknerBorkovec1990}%
  \BibitemOpen
  \bibfield  {author} {\bibinfo {author} {\bibfnamefont {P.}~\bibnamefont
  {H\"anggi}}, \bibinfo {author} {\bibfnamefont {P.}~\bibnamefont {Talkner}},\
  and\ \bibinfo {author} {\bibfnamefont {M.}~\bibnamefont {Borkovec}},\
  }\bibfield  {title} {\bibinfo {title} {Reaction-rate theory: fifty years
  after kramers},\ }\href {https://doi.org/10.1103/revmodphys.62.251}
  {\bibfield  {journal} {\bibinfo  {journal} {Reviews of Modern Physics}\
  }\textbf {\bibinfo {volume} {62}},\ \bibinfo {pages} {251} (\bibinfo {year}
  {1990})}\BibitemShut {NoStop}%
\bibitem [{\citenamefont {Hogenboom}\ \emph {et~al.}(1967)\citenamefont
  {Hogenboom}, \citenamefont {Webb},\ and\ \citenamefont
  {Dixon}}]{Hogenboom1967}%
  \BibitemOpen
  \bibfield  {author} {\bibinfo {author} {\bibfnamefont {D.~L.}\ \bibnamefont
  {Hogenboom}}, \bibinfo {author} {\bibfnamefont {W.}~\bibnamefont {Webb}},\
  and\ \bibinfo {author} {\bibfnamefont {J.~A.}\ \bibnamefont {Dixon}},\
  }\bibfield  {title} {\bibinfo {title} {{Viscosity of several liquid
  hydrocarbons as a function of temperature, pressure, and free volume}},\
  }\href {https://doi.org/10.1063/1.1841088} {\bibfield  {journal} {\bibinfo
  {journal} {J. Chem. Phys.}\ }\textbf {\bibinfo {volume} {46}},\ \bibinfo
  {pages} {2586} (\bibinfo {year} {1967})}\BibitemShut {NoStop}%
\bibitem [{\citenamefont {Naghizadeh}(1964)}]{Naghizadeh1964}%
  \BibitemOpen
  \bibfield  {author} {\bibinfo {author} {\bibfnamefont {J.}~\bibnamefont
  {Naghizadeh}},\ }\bibfield  {title} {\bibinfo {title} {{Diffusion and glass
  transition in simple liquids}},\ }\href {https://doi.org/10.1063/1.1713585}
  {\bibfield  {journal} {\bibinfo  {journal} {J. Appl. Phys.}\ }\textbf
  {\bibinfo {volume} {35}},\ \bibinfo {pages} {1162} (\bibinfo {year}
  {1964})}\BibitemShut {NoStop}%
\bibitem [{\citenamefont {Takagi}\ and\ \citenamefont
  {Negishi}(1980)}]{Takagi1980}%
  \BibitemOpen
  \bibfield  {author} {\bibinfo {author} {\bibfnamefont {K.}~\bibnamefont
  {Takagi}}\ and\ \bibinfo {author} {\bibfnamefont {K.}~\bibnamefont
  {Negishi}},\ }\bibfield  {title} {\bibinfo {title} {{Measurement of
  ultrasonic relaxation time and mean free path in liquids}},\ }\href
  {https://doi.org/10.1063/1.439298} {\bibfield  {journal} {\bibinfo  {journal}
  {J. Chem. Phys.}\ }\textbf {\bibinfo {volume} {72}},\ \bibinfo {pages} {1809}
  (\bibinfo {year} {1980})}\BibitemShut {NoStop}%
\bibitem [{\citenamefont {Rosenfeld}(1977)}]{Rosenfeld1977}%
  \BibitemOpen
  \bibfield  {author} {\bibinfo {author} {\bibfnamefont {Y.}~\bibnamefont
  {Rosenfeld}},\ }\bibfield  {title} {\bibinfo {title} {{Relation between the
  transport coefficients and the internal entropy of simple systems}},\ }\href
  {https://doi.org/10.1103/PhysRevA.15.2545} {\bibfield  {journal} {\bibinfo
  {journal} {Phys. Rev. A}\ }\textbf {\bibinfo {volume} {15}},\ \bibinfo
  {pages} {2545} (\bibinfo {year} {1977})}\BibitemShut {NoStop}%
\bibitem [{\citenamefont {Rosenfeld}(1999)}]{Rosenfeld1999}%
  \BibitemOpen
  \bibfield  {author} {\bibinfo {author} {\bibfnamefont {Y.}~\bibnamefont
  {Rosenfeld}},\ }\bibfield  {title} {\bibinfo {title} {{A quasi-universal
  scaling law for atomic transport in simple fluids}},\ }\href
  {https://doi.org/10.1088/0953-8984/11/28/303} {\bibfield  {journal} {\bibinfo
   {journal} {J. Phys. Condens. Matter}\ }\textbf {\bibinfo {volume} {11}},\
  \bibinfo {pages} {5415} (\bibinfo {year} {1999})}\BibitemShut {NoStop}%
\bibitem [{\citenamefont {Galliero}\ \emph {et~al.}(2011)\citenamefont
  {Galliero}, \citenamefont {Boned},\ and\ \citenamefont
  {Fernndez}}]{Galliero2011}%
  \BibitemOpen
  \bibfield  {author} {\bibinfo {author} {\bibfnamefont {G.}~\bibnamefont
  {Galliero}}, \bibinfo {author} {\bibfnamefont {C.}~\bibnamefont {Boned}},\
  and\ \bibinfo {author} {\bibfnamefont {J.}~\bibnamefont {Fernndez}},\
  }\bibfield  {title} {\bibinfo {title} {{Scaling of the viscosity of the
  Lennard-Jones chain fluid model, argon, and some normal alkanes}},\
  }\bibfield  {journal} {\bibinfo  {journal} {J. Chem. Phys.}\ }\textbf
  {\bibinfo {volume} {134}},\ \href {https://doi.org/10.1063/1.3553262}
  {10.1063/1.3553262} (\bibinfo {year} {2011})\BibitemShut {NoStop}%
\bibitem [{\citenamefont {Bell}\ \emph {et~al.}(2019)\citenamefont {Bell},
  \citenamefont {Messerly}, \citenamefont {Thol}, \citenamefont {Costigliola},\
  and\ \citenamefont {Dyre}}]{Bell2019}%
  \BibitemOpen
  \bibfield  {author} {\bibinfo {author} {\bibfnamefont {I.~H.}\ \bibnamefont
  {Bell}}, \bibinfo {author} {\bibfnamefont {R.}~\bibnamefont {Messerly}},
  \bibinfo {author} {\bibfnamefont {M.}~\bibnamefont {Thol}}, \bibinfo {author}
  {\bibfnamefont {L.}~\bibnamefont {Costigliola}},\ and\ \bibinfo {author}
  {\bibfnamefont {J.~C.}\ \bibnamefont {Dyre}},\ }\bibfield  {title} {\bibinfo
  {title} {{Modified Entropy Scaling of the Transport Properties of the
  Lennard-Jones Fluid}},\ }\href {https://doi.org/10.1021/acs.jpcb.9b05808}
  {\bibfield  {journal} {\bibinfo  {journal} {J. Phys. Chem. B}\ }\textbf
  {\bibinfo {volume} {123}},\ \bibinfo {pages} {6345} (\bibinfo {year}
  {2019})}\BibitemShut {NoStop}%
\bibitem [{\citenamefont {Viet}\ \emph {et~al.}(2022)\citenamefont {Viet},
  \citenamefont {Khennache}, \citenamefont {Galliero}, \citenamefont {Alapati},
  \citenamefont {Nguyen},\ and\ \citenamefont {Hoang}}]{Viet2022}%
  \BibitemOpen
  \bibfield  {author} {\bibinfo {author} {\bibfnamefont {T.~Q.~Q.}\
  \bibnamefont {Viet}}, \bibinfo {author} {\bibfnamefont {S.}~\bibnamefont
  {Khennache}}, \bibinfo {author} {\bibfnamefont {G.}~\bibnamefont {Galliero}},
  \bibinfo {author} {\bibfnamefont {S.}~\bibnamefont {Alapati}}, \bibinfo
  {author} {\bibfnamefont {P.~T.}\ \bibnamefont {Nguyen}},\ and\ \bibinfo
  {author} {\bibfnamefont {H.}~\bibnamefont {Hoang}},\ }\bibfield  {title}
  {\bibinfo {title} {{Mass Effect on Viscosity of Mixtures in Entropy Scaling
  Framework: Application to Lennard-Jones mixtures}},\ }\href
  {https://doi.org/10.1016/j.fluid.2022.113459} {\bibfield  {journal} {\bibinfo
   {journal} {Fluid Phase Equilib.}\ ,\ \bibinfo {pages} {113459}} (\bibinfo
  {year} {2022})}\BibitemShut {NoStop}%
\bibitem [{\citenamefont {Tarjus}\ \emph {et~al.}(2004)\citenamefont {Tarjus},
  \citenamefont {Kivelson}, \citenamefont {Mossa},\ and\ \citenamefont
  {Alba-Simionesco}}]{Tarjus2004}%
  \BibitemOpen
  \bibfield  {author} {\bibinfo {author} {\bibfnamefont {G.}~\bibnamefont
  {Tarjus}}, \bibinfo {author} {\bibfnamefont {D.}~\bibnamefont {Kivelson}},
  \bibinfo {author} {\bibfnamefont {S.}~\bibnamefont {Mossa}},\ and\ \bibinfo
  {author} {\bibfnamefont {C.}~\bibnamefont {Alba-Simionesco}},\ }\bibfield
  {title} {\bibinfo {title} {{Disentangling density and temperature effects in
  the viscous slowing down of glassforming liquids}},\ }\href
  {https://doi.org/10.1063/1.1649732} {\bibfield  {journal} {\bibinfo
  {journal} {J. Chem. Phys.}\ }\textbf {\bibinfo {volume} {120}},\ \bibinfo
  {pages} {6135} (\bibinfo {year} {2004})}\BibitemShut {NoStop}%
\bibitem [{\citenamefont {Hentschel}\ \emph {et~al.}(2012)\citenamefont
  {Hentschel}, \citenamefont {Karmakar}, \citenamefont {Procaccia},\ and\
  \citenamefont {Zylberg}}]{Hentschel2012}%
  \BibitemOpen
  \bibfield  {author} {\bibinfo {author} {\bibfnamefont {H.~G.~E.}\
  \bibnamefont {Hentschel}}, \bibinfo {author} {\bibfnamefont {S.}~\bibnamefont
  {Karmakar}}, \bibinfo {author} {\bibfnamefont {I.}~\bibnamefont
  {Procaccia}},\ and\ \bibinfo {author} {\bibfnamefont {J.}~\bibnamefont
  {Zylberg}},\ }\bibfield  {title} {\bibinfo {title} {{Relaxation mechanisms in
  glassy dynamics: The Arrhenius and fragile regimes}},\ }\href
  {https://doi.org/10.1103/PhysRevE.85.061501} {\bibfield  {journal} {\bibinfo
  {journal} {Phys. Rev. E - Stat. Nonlinear, Soft Matter Phys.}\ }\textbf
  {\bibinfo {volume} {85}},\ \bibinfo {pages} {1} (\bibinfo {year} {2012})},\
  \Eprint {https://arxiv.org/abs/1202.1127} {arXiv:1202.1127} \BibitemShut
  {NoStop}%
\bibitem [{\citenamefont {Parmar}\ \emph {et~al.}(2020)\citenamefont {Parmar},
  \citenamefont {Guiselin},\ and\ \citenamefont {Berthier}}]{parmar2020stable}%
  \BibitemOpen
  \bibfield  {author} {\bibinfo {author} {\bibfnamefont {A.~D.}\ \bibnamefont
  {Parmar}}, \bibinfo {author} {\bibfnamefont {B.}~\bibnamefont {Guiselin}},\
  and\ \bibinfo {author} {\bibfnamefont {L.}~\bibnamefont {Berthier}},\
  }\bibfield  {title} {\bibinfo {title} {Stable glassy configurations of the
  kob--andersen model using swap monte carlo},\ }\href@noop {} {\bibfield
  {journal} {\bibinfo  {journal} {The Journal of Chemical Physics}\ }\textbf
  {\bibinfo {volume} {153}},\ \bibinfo {pages} {134505} (\bibinfo {year}
  {2020})}\BibitemShut {NoStop}%
\bibitem [{\citenamefont {Macedo}\ and\ \citenamefont
  {Litovitz}(1965)}]{MacedoLitovitz1965}%
  \BibitemOpen
  \bibfield  {author} {\bibinfo {author} {\bibfnamefont {P.~B.}\ \bibnamefont
  {Macedo}}\ and\ \bibinfo {author} {\bibfnamefont {T.~A.}\ \bibnamefont
  {Litovitz}},\ }\bibfield  {title} {\bibinfo {title} {On the relative roles of
  free volume and activation energy in the viscosity of liquids},\ }\href
  {https://doi.org/10.1063/1.1695683} {\bibfield  {journal} {\bibinfo
  {journal} {The Journal of Chemical Physics}\ }\textbf {\bibinfo {volume}
  {42}},\ \bibinfo {pages} {245} (\bibinfo {year} {1965})}\BibitemShut
  {NoStop}%
\end{thebibliography}
